\documentstyle[epsf]{mn}

\title{Non--LTE Excitation of H$_2$ in Magnetised Molecular Shocks}

\author[Ivan O'Brien, L O'C Drury]{Ivan O'Brien\thanks{e-mail: io@cp.dias.ie},
L O'C Drury \\
Dublin Institute for Advanced Studies, 5 Merrion Square, Dublin 2, Ireland}

\begin{document}

\maketitle

\begin{abstract}
The observed H$_2$ line ratios in OMC--1 and IC443 are not
satisfactorily explained by conventional shock excitation models.  We
consider the microscopic collisional processes implicit in ambipolar
diffusion models of magnetised C-shocks and show that non-LTE level
populations and emission line ratios are an inevitable consequence of
such models.  This has important implications for the use of molecular
hydrogen lines as diagnostics of shock models in molecular clouds.
\end{abstract}


\section{Introduction}
Recent observations of shocked molecular hydrogen, particularly in
OMC--1 and IC--443 (Brand {\it et al.} 1988, Richter, Graham \& Watt
1995), show H$_2$ excitation line ratios that have proven difficult to
model. The original models of these regions involved magnetic shocks,
and were able to explain the presence of relatively high velocity
shocks (over 25 km s$^{-1}$) which did not lead to molecular
destruction (Draine 1980). These models, however, could not explain
the observed line ratios. New models for these regions suggest that
the observed populations could result from the complex cooling zones
which follow partially--dissociative shocks (Brand {\em et al.}  1988)
or result from emission in the wake of fast--moving clumps of material
(Smith 1991). However, no model has yet been able to explain the
uniformity of the emission over large regions and why it appears to be
so similar for two different sources. It is possible that a magnetic
shock model which incorporated non--thermal effects could help to
explain both the line widths and ratios observed.

Draine (1980) and Draine, Roberge \& Dalgarno (1983, hereafter DRD83)
studied magnetohydrodynamic (MHD) shock models, whereby the shock
structure is altered by the streaming of ionised species ahead of the
neutral shock front due to magnetic field compression. Both papers
note, however, that the low densities present ($<10^{6}$cm$^{-3}$)
mean that it is uncertain if LTE can be assumed throughout these
shocks. This uncertainty is strongest if particles have anomalous
excitation populations, something which will happen whenever ions
collide with neutrals at highly non--thermal velocities, an integral
part of the differential streaming (ambipolar diffusion) process at
the heart of MHD shocks. In such a collision, and in the collisional
cascade which follows, there may be significant non--thermal
excitation of the neutral particles, leading to enhanced emission from
lines not normally seen at the local kinetic temperature. In this
paper we describe detailed Monte--Carlo simulations of the microscopic
processes of momentum transfer between the different component species
in the shock to see what effect such non--LTE processes have on the
total emission and the relative line intensity ratios emitted from a
magnetised shock. These results are compared with observation, and
found to have many of the same properties without resort to complex
geometry.

\section{The Model}

In the shocked regions of interest H$_2$ is the principal coolant
(Smith, 1991). Since it is the ratios of H$_2$ emission lines that are
of interest and other species have relative abundances of less than
10$^{-4}$, it is reasonable to limit the chemical composition of the
neutral gas considered to just this one species. More complex
chemistry could, in principal, be included but is unlikely to affect the
results significantly.

In ambipolar diffusion models of shocks (Draine 1980, DRD83 for
example. See Draine \& McKee 1993 for a review) the small number of
ions present are forced at relatively high velocity through the bulk
neutral gas by electromagnetic forces associated with the compression
of the magnetic field in the shock. At the low ionisation fractions in
these regions, typically $10^{-4}-10^{-8}$, reasonable assumptions
about the local ambient magnetic field show that the ``streaming
velocity'' can be a large fraction of the shock velocity, reaching 20
km s$^{-1}$ for a 25 km s$^{-1}$ shock, for example. Collisions
between the fast moving ions and the neutral molecules transfer
momentum and energy to the neutral population thereby accelerating and
heating it.  In all previous calculations it has been assumed that the
neutral population is in LTE, so that the radiation resulting from
this heating can be calculated using standard cooling functions and
line ratios.  However the inital ion-neutral collision is at
velocities which are typically very much higher than thermal, and the
resulting accelerated neutral molecule will also be very fast moving,
as will the second generation of molecules emerging from its next
collision, and so on. There will be a cascade of collisions with the
velocity roughly halving at each generation, and the number of
molecules involved doubling. At least this would be the picture if
there was no excitation of internal degrees of freedom in the
collisions. The Monte--Carlo simulations described in this paper model
collisional cascades in molecular hydrogen using the best available
data for the collisional excitation and deexcitation of the various
rotational and vibrational levels.

\section{Approximations \& Assumptions}

The ions are taken to be ``typical'' with a single charge unit and
mass ${\rm m}_{\rm i}=30 {\rm m}_{\rm H}$ (e.g.\ CO$^+$) following de
Jong, Dalgarno \& Boland (1980). No excitation of the neutrals will be
considered in ion--neutral collisions, due to uncertainty in the
transition rates and the likelihood of a strong dependence on the
exact ion species. As these are the highest energy collisions and will
probably lead to significant excitation this will reduce the
excitation level resulting from the cascade.  The only interaction
allowed, therefore, is momentum transfer, with differential
cross--sections given by an isotropic post--collision momentum
distribution in the centre--of--mass (CM) frame. This is insensitive
to the mass of the ion as long as it is significantly heavier than the
neutral. The ion--neutral collision velocity is the vector sum of the
neutral velocity, the ion--neutral streaming velocity, the random
velocity of the ion along the magnetic field line and the velocity
around this line. This is, on average, slightly greater than the
streaming velocity, but can be up to twice this value.

Electrons can be important in (de)excitation of hydrogen molecules
when they are sufficiently energetic. The additional kinetic energy
gained by the electrons due to magnetic field compression is, however,
small compared to their thermal kinetic energy and so these
interactions will not be important compared to the high--energy
ion--neutral ones. In this model they will be ignored.

The kinematic properties of grains in magnetised regions are highly
uncertain. If they are tightly coupled to the magnetic field then
their inertia leads to a large reduction in the ion--magnetosonic
velocity, suppressing C--type shocks (McKee {\em et al.} 1984). We
will assume that this is not the case.

The data available for the main interaction process, H$_2$--H$_2$
collisions, are very sparse. Some experiments have been carried out to
determine the state--specific differential cross--sections, but only
at a single energy and using deuterised species (Buck {\it et al.}
1983a,b) which have different internal energy properties. This data
is, therefore, unusable in this calculation. Instead, an isotropic
post--collision momentum distribution in the CM will be used, as it
distributes momentum rapidly and so minimises the longevity of
non--LTE effects. It is expected that for elastic collisions, where
the interaction is weaker, the momentum transfer will be lower than
this. When data for these processes become available they can be added
to the model.

\begin{figure}
\begin{center}
\leavevmode
\epsfxsize=240pt \epsfbox{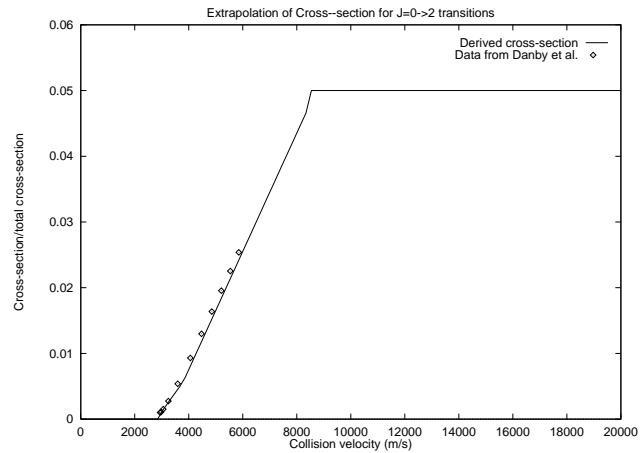}
\caption{Typical extrapolation used for fits to cross--sections from 
Danby {\it et al.} (1987) up to higher collision velocities. The rates
were linearly extrapolated beyond the available data and then cut off.
This cutoff is unlikely to take place, particularly for high--energy
transitions.}
\label{fig:extrapolate}
\end{center}
\end{figure}

Internal excitation and deexcitation cross--sections are also poorly
known. Various extrapolations have been published (DRD83, Lepp \&
Schull 1983) of the few experiments which have been carried out (Dove
\& Teitelbaum 1974) for combined V, J transitions. The rates published
by Lepp \& Schull covered combined transitions with
J$\rightarrow$J,J$\pm$2. These will be used and trivially extended to
cover transitions with J$\rightarrow$J$\pm$4,J$\pm$6, but no further,
following DRD83, to avoid gross extrapolation errors. Rates for pure J
transitions with V=0 are better known, though still only for a limited
range of transitions and energies. Danby, Flower \& Montiero (1987)
give the most reliable rates, and the unintegrated cross--sections
used to generate these rates (Flower, private communication) have been
used and extrapolated from the original data for J$\rightarrow$J$\pm$2
for J up to 6. These extrapolations are for higher energy states than
those considered, transitions up to J$\rightarrow$J$\pm$6 and for
higher energies than those calculated. The internal state of the
colliding particle is not considered important, as the available data
(Danby {\em et al.} 1987) only considers partners in the ground state. All of
the rate extrapolations carried out have been extremely conservative
(see figure~\ref{fig:extrapolate} for an example) to ensure that any
non--thermal effects seen can be considered lower bounds.

Conversions between thermal collision rates and velocity
cross--sections were carried out by application of a resolving kernel
and microscopic reversibility was used to transform between upward and
downward cross--sections (O'Brien 1995).

\section{Implementation}

\begin{figure}
\begin{center}
\leavevmode
\epsfxsize=240pt \epsfbox{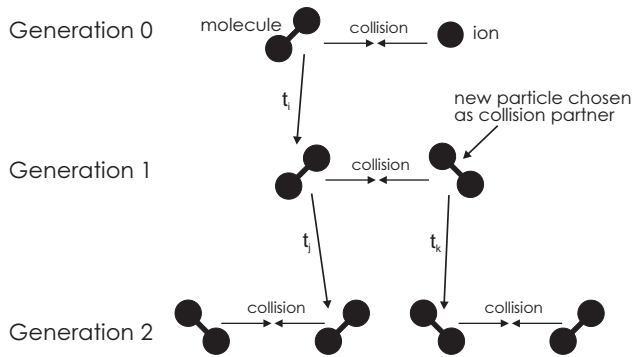}
\caption{The collisional cascade system. Each ``generation'' involves 
a single collision for all particles in the system. The collision
partners for these particles are selected statistically, as described
in the text. The t$_i$s are the times between collisions.}
\label{fig:generation}
\end{center}
\end{figure}

A collisional cascade is considered to start with the initial
neutral--neutral collision following an ion--neutral collision and
continues until the kinetic energies of the particles are
thermalised. With the differential cross--sections used this takes
place after about 10 ``generations''. The concept of a collisional
generation is illustrated in figure~\ref{fig:generation}. In each
collision, in addition to the transfer of momentum, there is a chance
that one, or both, of the molecules will become internally excitated
or deexcited. Between collisions these excited molecules can
radiate.

For practical reasons these simulations must avoid having to track
large numbers of particles through space with a precision great enough
to determine collisional impact parameters for molecular
collisions. Thus the only information which can be used to determine
collisions is the relative velocities of all the molecules and the
system density. A total collisional cross--section allows selection of
collision partners, as detailed below, while partial cross--sections
for deflection angle and internal (de)excitation allow the outcome of
the resulting collisions to be determined.

Collision times and partners are selected by the following
process. Consider two particles, $i$ and $j$, in a box with repeating
boundary conditions. The average time for particle $i$ to collide with
particle $j$ is given by $t_{ij} = 1/\rho_j \sigma v_{ij}$ where
$\rho_j$ is the number density of particle $j$, $\sigma$ is the total
collisional cross--section and $v_{ij}$ is the relative velocity of
$i$ and $j$. The time to an actual collision is given by
$\tau_{ij}=t_{ij}R$, where $R$ is an exponentially distributed random
variable centered on 1. If there are many particles in the box then
the time until the next collision that particle $i$ undergoes,
$\tau_i$, is $\min_j \tau_{ij}$. In a full system this minimisation is
taken over 10 potential collision partners selected from a population
with the correct velocity distribution (taken to be Maxwellian at a
specified temperature). For the purposes of calculating the times each
is taken to have a number density on tenth that of the local medium.
This procedure has been shown to give the correct relative velocity
distribution function and collision time distribution over a
sufficient number of collisions for a non--radiating, thermal gas, the
only system for which such distributions are well--known.

When a collision partner has been selected its internal state is
calculated. For low--temperature systems the internal states are taken
to be initially thermal at the kinetic temperature. At higher
temperatures this may not be a good assumption, due to depletion in
any unexcited gas due to long collisional reexcitation times (Chang \&
Martin 1991, for example) or residual excitation from previous
high--energy collisional cascades. The initial population is not
critical, however, as the non--thermal populations are much higher for
ion velocities in excess of 10 kms$^{-1}$. In simulations of entire
shocks the evolution of these states may be important. An ortho--para
ratio of 3 is used throughout, and there are no ortho--para conversion
mechanisms included in the simulation.

In order to determine the effects of heating a large body of gas over
a wide temperature range a series of cascades is necessary,
incorporated into a complete dynamical shock model. While this is not
yet possible using this model the likely properties of such a system
are indicated in the short chains of cascades that have been
calculated.

\section{Results}
\begin{figure}
\begin{center}
\leavevmode
\epsfxsize=240pt \epsfbox{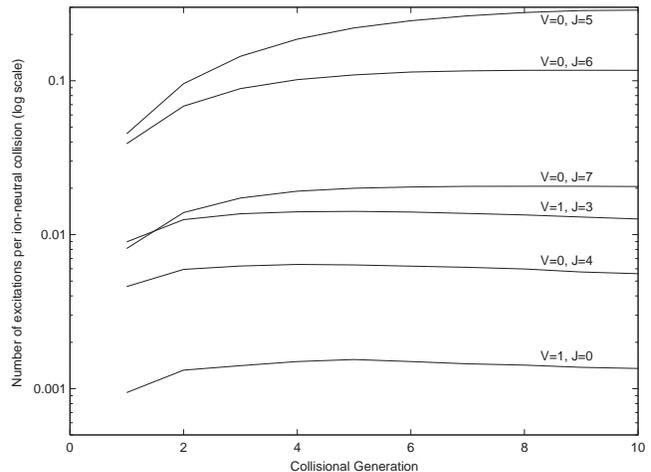}
\caption{The evolution of population with cascade generation for a 
selection of internal states for a system with T=500K, $\rho=10^4 {\rm
cm}^{-3}$ and v$_{\rm in}=15 {\rm km s}^{-1}$, typical of the early
stages of a MHD shock. There are no excitations in the 0th generation,
the ion--neutral collision. Initial collisional excitation of high
energy states is followed by collisional and radiative depopulation,
though at these densities a large proportion of the excitations are
still present after 10 generations.}
\label{fig:pops}
\end{center}
\end{figure}

There is a strong non--thermal character both to the radiation and
residual excitation resulting from collisional cascades.  Rapid
excitation to high--energy internal states in the first few
collisional generations, when the non--thermal energies are highest,
is followed by gradual deexcitation due both to collisional and
spontaneous deexcitation. The population changes over time are
indicated in figure~\ref{fig:pops}. The shorter radiative deexcitation
times of, and existence of multiple decay paths for, more highly
excited states means that they decay more rapidly, although the
logarithmic scale on the y--axis tends to disguise this. As the
excitation energy for H$_2$ is so high (the lowest excitation is at
510K) most molecules will initially be in the ground state of ortho--
or para--H$_2$ (V=0, J=1,0) at the temperatures considered. Combined
with the excitation cutoff at $J\rightarrow J+6$, this means that the
populations of states with $J>7$ are lower than their energy would
suggest, as they require multiple collisions to become
populated. However, even within a ten generation cascade a significant
number of particles can undergo multiple collisional excitations.

\begin{figure}
\begin{center}
\leavevmode
\epsfxsize=240pt \epsfbox{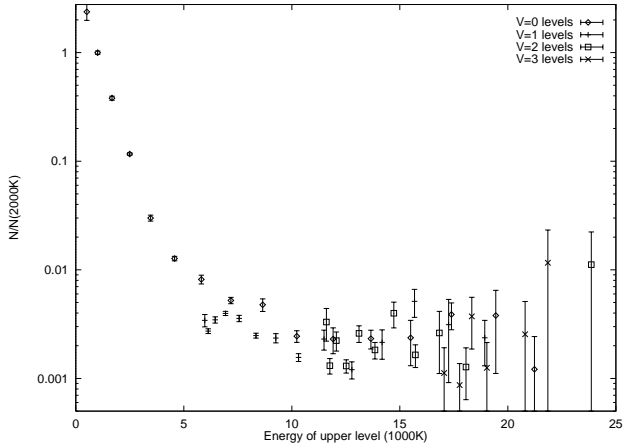}
\caption{The relative populations of internal states of H$_2$, as 
derived from the cascade radiation, for the system in figure 3. The
populations are shown relative to those expected for a thermal system
at 2000K, following Brand {\it et al.} 1988. Different V sequences are
differentiated to show that the energy of the level is the main factor
determining the relative population, in line with observations. Error
bars are formal only.}
\label{fig:relpop}
\end{center}
\end{figure}

The most important result of these simulations is the relative
populations of the various internal states of H$_2$ as derived from
radiation from the system. Figure~\ref{fig:relpop} shows these
populations for a system with parameters typical of the ambipolar
region of a MHD shock, relative to the expected populations for a
thermal gas at 2000K. This is so that the plot may be directly
compared with those in, for example, Brand {\it et al.} 1988 and
Richter {\it et al.}  1995. The population is plotted against the
energy of the upper level, in 1000K, and a logarithmic scale is used
on the y axis. The absolute scale on the y--axis will vary with the
proportion of molecules involved in cascades. This graph assumes that
all particles are in cascades, corresponding to an ionisation fraction
of roughly 10$^{-6}$.

Any straight line on this plot would indicate a gas at a single
temperature, so it is clear that no single temperature can explain all
these data, in marked contrast with previous magnetic shock
models. Instead, the excitation temperature of the internal states
shows a strong trend to increase with increasing level energy,
agreement with observed H$_2$ shocks in OMC--1 and IC443, previously
explained using complex multi--component shocks or cooling zones
following partially--dissociative shocks. The errors shown are
$\sqrt{N}$ only. As the absolute relative populations of the higher
energy states are very low, albeit much higher than for thermal
systems, reducing the errors in these populations is very expensive
computationally.

The shape of the curve in figure~\ref{fig:relpop} is a function of the
neutral kinetic temperature, ion--neutral streaming velocity (v$_{\rm
in}$) and neutral particle density, as described below.

\begin{figure}
\begin{center}
\leavevmode
\epsfxsize=240pt \epsfbox{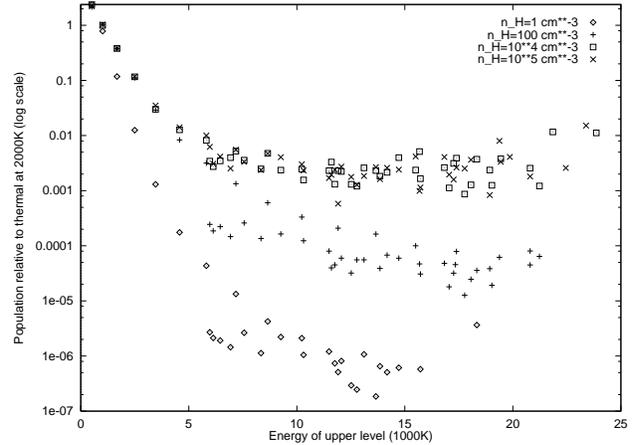}
\caption{Derived level populations against level energy for a range 
of densities for a system at 500K with v$_{\rm in}$=15kms$^{-1}$. The
emission appears to be independent of density above $\rho=10^4 {\rm
cm}^{-3}$.}
\label{fig:rhocrit}
\end{center}
\end{figure}

\begin{itemize}

\item{\bf Neutral Kinetic Temperature:}
As the initial neutral internal populations are taken to be those for
a thermal system the temperature determines the emission from the
thermal lines, and has a small effect on those states which are not
thermally populated.

\item{\bf Ion--Neutral Streaming Velocity:}
As this determines the amount of non--thermal energy available in a
given cascade it also governs the number of non--thermal excitations
which occur and their energy range. Below v$_{\rm in} \simeq 10 {\rm
km s}^{-1}$ there are negligable levels of internal excitation, but
above this they increase strongly with v$_{\rm in}$.

\item{\bf Density:}
At low densities most excitations will lead to radiation whereas
collisional deexcitation becomes dominant at higher densities. There
appears to be a critical density above which the radiation from a
cascade is essentially invariant for a few orders of magnitude in
density, as shown in figure~\ref{fig:rhocrit}. This critical density
is just below the densities predicted for the H$_2$ shock in OMC--1
(Brand {\it et al.}  1988).

\end{itemize}

\begin{figure}
\begin{center}
\leavevmode
\epsfxsize=240pt \epsfbox{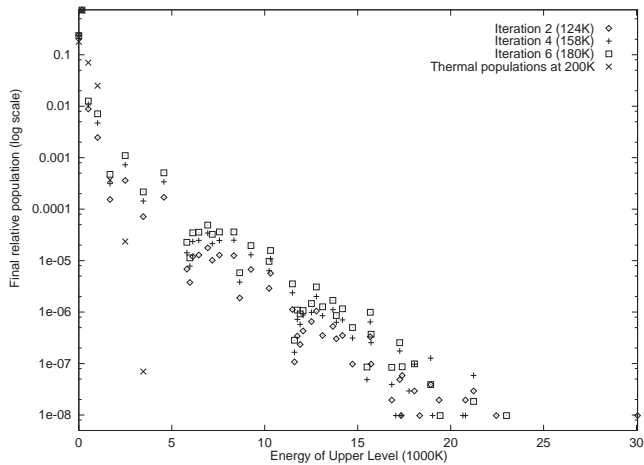}
\caption{Residual internal states after cascade simulations. Here the 
final populations and kinetic temperature from iteration 1 are used as
the inputs for iteration 2, and so on. v$_{\rm in}$=15 kms$^{-1}$,
$\rho$=10$^{4}$cm$^{-3}$ in all runs. There is a definate buildup of
populations of higher energy states, well in excess of the
corresponding increase in kinetic temperature.}
\label{fig:manyruns}
\end{center}
\end{figure}

Using these calculations to model an entire shock is not yet possible,
chiefly due to computational constraints. The effects of residual
enhancements in population levels following cascades can be seen,
however, in calculations where these populations are used as the input
to subsequent cascades. After a series of such cascades it is clear
that this enhancement does occur, even for relatively low densities,
as illustrated in figure~\ref{fig:manyruns}. Due to these increased
internal populations later runs also have enhanced radiation from
higher energy states. At each stage the kinetic temperature increases
as indicated, while the rotational and vibrational populations
increase more rapidly. The effect is seen most strongly at higher
energies, where the residual populations are several orders of
magnitude higher than would be seen in a thermal gas. These states
cannot easily be excited from the ground state, and the residual
excitation populations enable them to be populated.

The observed internal populations, which are an unavoidable result of
the process of ambipolar diffusion, are strongly reminiscent of those
which have been observed in H$_2$ shocks in OMC--1 and IC443, where
the excitation temperature is also an increasing function of level
energy and is almost independent of the details of V and J
state. There are two major differences between the data sets. The
first is the absolute magnitude of the radiation, which is higher for
observed systems than in the data presented here. For observations
this scale will depend on assumptions about the system density. For
this model it will vary with the ratio of ion--neutral collisions to
pure neutral ones and, therefore, with the ionisation fraction. The
scales shown assume an ionisation fraction of approximately
10$^{-6}$. The second major difference is the behaviour of the
populations above about 15000K. These have been observed by Brand {\em
et al.} (1988) and Richter {\em et al.} (1995) to rise dramatically
whereas these calculations show, at best, a modest increase. This is
probably due to the conservative nature of the extrapolations and the
arbitrary excitation cutoff at J$\rightarrow$J+6, both of which limit
the populations of highly excited states, as well as the differential
cross--sections used, which disperse energy extremely efficiently and
thus serve to reduce the population of very high velocity particles
rapidly.

\section{Further Work Needed}
The main difficulty with applying these calculations is the uncertain
and unreliable nature of both differential and internal excitation
rate coefficients. This work shows the importance of high--quality
quantum mechanical calculations of these coefficients if H$_2$ line
ratios are to be used as diagnostics in molecular clouds.

A calculation of the integrated emission from a complete shock is
needed to test the importance of these results both for the dynamics
of shocks and the resulting molecular line ratios. Such a calculation
will be computationally expensive. Simple order--of--magnitude
calculations show that the line strengths in figure~\ref{fig:relpop}
correspond to those which would be observed from a magnetic shock
travelling at 25 km s$^{-1}$ through a medium with $\rho \sim 10^4
{\rm cm}^{-3}$ and $10^{-7} \leq {\rm x}_i \leq 10^{-4}$, typical
conditions in these regions. The first few ($<$1000K) states will also
have a significant contribution from thermal emission for ionisation
fractions of this order.

\section{Conclusions}
Non--LTE excitation populations in H$_2$ are a natural consequence of
ambipolar diffusion in molecular shocks, with no need for complex
geometrical constructions or multicomponent shock models. Comparison
of figure~\ref{fig:relpop} with observations in Brand {\em et al.}
(1988), for example, show that this mechanism could explain the
observed line ratios in OMC--1 and IC443, with the reservation that
the poor quality of available cross--sections does not allow firm
predictions to be made. The conservative nature of these calculations
mean that the upturn in excitation populations which have been seen
observationally at higher energies could result from this process
also. Full shock models incorporating this theory are possible and
should be carried out to determine the integrated emission over whole
shock regions.

\section*{Acknowledgements}
IOB would like to thank Dr Alan Moorehouse for the conversations which
led to this work, and Drs B.T. Draine and P.W.J.L. Brand for helpful
discussions. This paper comprises part of the research contained in
IOBs PhD thesis (O'Brien 1995), which was supervised by Dr. S. McMurry
of Trinity College, Dublin. This work was part funded by FORBAIRT
grant number BR/91/139.

\section*{References}
\newcommand{\rf}{\par\noindent\hangindent 20pt}

\rf{Brand, P.W.J.L., Moorhouse, A., Burton, M.G., Geballe, T.R., Bird, M. and
Wade, R., 1988, {\it Ap. J.}, {\bf 334}, L103}

\rf{Buck, U., Huisken, F., Kohlhase, A., Otten, D., 1983a, {\it J. Chem. Phys},
{\bf 78}, 4439}

\rf{Buck, U., Huisken, F., Maneke, G., 1983b, {\it J. Chem. Phys}, {\bf 78},
4430}

\rf{Chang, C.A., Martin, P.G., 1991, {\it Ap. J.}, {\bf 378}, 202}

\rf{Danby, G., Flower, D.R., Montiero, T.S., 1987 {\it M.N.R.A.S.}, {\bf 226},739}

\rf{de Jong, T., Dalgarno, A., Boland, W., 1980, {\it Astr. Ap.}. {\bf 91}, 68}

\rf{Dove, J.E. and Teitelbaum, H., 1974, {\it Chem. Phys.}, {\bf 6}, 431}

\rf{Draine, B.T., 1980, {\it Ap. J.}, {\bf 241}, 1021}

\rf{Draine, B.T., McKee, C.F., 1993, {\it Ann. Rev. Astron. Astrophys.}, 
373-432}

\rf{Draine, B.T., Roberge, W.G., Dalgarno, A., 1983, {\it Ap. J.}, 
{\bf 264}, 485}

\rf{Lepp, S., Schull, J.M., 1983, {\it Ap. J.}, {\bf 270}, 578}

\rf{McKee, C.F., Chernoff, D.F., Hollenbach, D.J., 1984, 
in {\it Galactic and Extragalactic Infrared Spectroscopy}, 
ed. Kessler, M.F., Phillips, J.P., 103, Dordrecht: Reidel}
\rf{Richter, M.J., Graham, J.R., Wright, 1995a, {\it Ap. J.}, submitted}

\rf{O'Brien, I.T., 1995, PhD Thesis}

\rf{Smith, M.D., 1991, {\it M.N.R.A.S.}, {\bf 253}, 175}

\end{document}